\documentclass{article}
\usepackage{spconf,amsmath,algpseudocode,algorithm,comment}

\title{IMPROVED LOSSLESS CODING FOR STORAGE AND TRANSMISSION OF MULTICHANNEL IMMERSIVE AUDIO} 
\name{Toni Hirvonen, Mahmoud Namazi}
\address{Samsung Research America}
\begin{document}

© 2023 IEEE. Personal use of this material is permitted. Permission from IEEE must be obtained for all other uses, in any current or future media, including reprinting/republishing this material for advertising or promotional purposes, creating new collective works, for resale or redistribution to servers or lists, or reuse of any copyrighted component of this work in other works.

\newpage

%
\maketitle
\begin{abstract}
In this paper, techniques for improving multichannel lossless coding are examined.
A method is proposed for the simultaneous coding of two or more different renderings (mixes) of the same content. 
The signal model uses both past samples of the upmix, and the current time samples of downmix samples to predict the upmix. 
Model parameters are optimized via a general linear solver, and the prediction residual is Rice coded. 
Additionally, the use of an SVD projection prior to residual coding is proposed. 
A comparison is made against various baselines, including FLAC. 
The proposed methods show improved compression ratios for the storage and transmission of immersive audio.

\end{abstract}
\section{Introduction}
\label{sec:intro}

As immersive audio gains popularity and content creators utilize the capabilities of these modern formats, 
the same content is increasingly available for different multichannel listening setups, such as 5.1 and 7.1.4  \cite{AC-4, MPEG-H}. 
Legacy content is also being transferred to these new formats.
To facilitate this, the different versions have to be created manually, or via an automatic process of re-mixing or rendering. 
While downmixing and rendering are mainly simple linear processes, 
upmixers have to reconstruct unknown signal components with more 
complicated algorithms. In the case of blind upmixing especially, the 
original artistic intent is not guaranteed to be preserved despite the 
sophistication of such systems, making them less than ideal.

Non-blind upmixing on the other hand can be viewed as being synonymous to
audio coding. Multichannel audio transmission and storage typically utilizes 
parametric coding, 
e.g. preserving the channel covariance 
structure is effective
\cite{AC-4, MPEG-H}. 
Unfortunately, lossy multichannel coding is difficult to optimize perceptually. 
Perceptual differences are difficult to 
judge due to their multidimensional nature \cite{zacharovspatial}.
While many codecs make good arguments that they achieve transparency 
after some bitrate, this cannot be fully guaranteed for all possible content 
due to the limitations of subjective testing.

Lossless coding is a viable option to address concerns related to 
both blind upmixing and parametric coding. 
As transmission capacities have constantly improved, 
the need for extremely low bitrates is no longer as major a concern as before.
Furthermore, prejudices against lossy coding and transmission have 
increased as both consumers and content creators become more educated. 
There is a need for exact control 
of the immersive audio reproduction process in all situations.

Compared to traditional lossless coding, it is not as clear how to most 
effectively deal with immersive audio.
More sophisticated prediction models have been proposed  
in \cite{MPEG-ALS, ambilossless}. 
However, in the case of multichannel audio, these models have 
not resulted in major benefits, 
but rather in small improvements.
As we see in Sec.\ \ref{sec:methods}, a system using a very simple baseline model of coding channels separately 
is able to get very close to real codec performance. 

This paper proposes methods to move toward more comprehensive handling of immersive lossless audio.
Our main contribution is to 
propose a hypothetical audio system, where several 
(two or more) different mixes are stored simultaneously for the same content.
Such a system would be possible to implement by packing the differently 
coded bitstreams in the same file container etc. 
Alternatively, the downmix(es) can be assumed to be available a priori at the 
decoder. However, having mixes in the same container can very effectively 
control the artistic intent for the content as well, regardless of compression. 

We construct a controlled experiment showing the attainable benefits 
of using \emph{hierarchical} reconstruction of the different formats.
The method exploits correlations between the different content versions and 
reconstructs more elaborate presentations based on the lower-level 
multichannel formats and a non-trivial signal model. 
This would then result in decreased storage requirements for the 
audio format described above.

In the use case of streaming a single format at a time, 
we additionally propose a method combining short-term prediction, SVD, and Rice 
coding which performs considerably better than the realistic baselines for 5.0 audio, 
at a computational encoding cost. 
Details of the methods are 
presented in Sec.\ \ref{sec:methods}. The experiment results, and 
discussion about their implications follow in Sec.\ \ref{sec:experiments} and 
Sec.\ \ref{sec:discussion}.

\section{METHODS}
\label{sec:methods}

\subsection{Core lossless coding engine}
\label{sec:engine}

Current real-life lossless codecs share many compression principles, 
techniques, as well as overall performance.
FLAC (Free Lossless Audio Codec) \cite{flac} is an open source, widely-used 
codec implementation whose fundamental algorithms are based on earlier
Shorten \cite{shorten}.
In this paper, we apply the official FLAC implementation as a reference, and replicate its 
performance with a simple baseline implementation.

The general principle applied in FLAC and our method can be described 
with the following simplified signal model (time and channel indices omitted):

\begin{equation}
    s = f(s') + e, 
\end{equation}

\noindent where the original signal $s$ is represented by 
a \emph{predictor function} $f()$
operating on a predictor source signal $s'$, which is 
often related to $s$. The prediction 
residual is noted with $e$. In standard lossless coding, $f()$
is often a linear predictor (LPC) operated on short frames, giving 
the signal model:

\begin{equation}\label{model:sep}
    s(t) = \sum_{k=1}^p{\beta_k s(t-k)} + e(t).
\end{equation}

\noindent For each time sample of the frame, $p$ 
(i.e. prediction order) past samples
are used as linear combination to model it. The coefficients 
$\beta = [\beta_1 ... \beta_p]$ are typically 
solved for minimizing the frame residual MSE $||e||^2_2$.
In real codecs, search procedures and frame signaling 
are often used to find the best $p$, as well as sometimes 
the type of predictor solution (e.g. LPC or a standard 
template \cite{shorten}). In this paper, we omit this optimization 
step, and rather aim to isolate the affect of the predictor source $s'$.

The compression ratio achieved by lossless audio codecs is 
predominantly effected by the entropy coding of the prediction residual $e$.
FLAC and Shorten simply assume that the residual distribution is 
geometric, with symmetrical focus around value zero.
These will have some Golomb code \cite{colomb} 
as an optimal prefix code.
Rice coding is a subset of such codes where the Golomb 
parameter is power of 2 for computational efficiency.

Without loss of data type generality, the length of the Rice 
code for an integer number $n$ can be obtained by 
Algorithm \ref{alg:colomb}. It calculates the codeword length 
in bits as a function of the Rice parameter $r$. 
Bitwise operations are utilized to 
perform sign-folding to non-negative integers, so that larger 
absolute values get longer codewords. Overflow checks are omitted here 
but may be implemented with $min$ operations.
The optimal Rice parameter $r$ can be estimated 
from the signal \cite{shorten}, but we used a 
brute-force search  
(e.g. $r \in [0...20]$) and selected the
code that minimizes the sum amount of bits in the
analysis frame of each channel.

\begin{algorithm}
\caption{Rice code length $l$ for integer $n$, given $r$}\label{alg:colomb}
\begin{algorithmic}
\Require $r \in [0, 1...]$
\State $n \gets int_{32}(n)$
\State $x \gets n \ll 1$
\State $y \gets n \gg 31$
\State $z \gets uint_{32}(\mathbin{x|y}) \gg r$
\State $l \gets 1 + r + z$
\end{algorithmic}
\noindent where $\ll$ indicates left-shift, $\gg$ 
right-shift, and $|$ XOR operation, all bitwise.
\end{algorithm}

Despite the simplicity of this baseline signal model, it already
accounts for much of the performance of the current real-life 
lossless codecs 
(see \ref{sec:results}). 
Some further tools, such as efficient handling of 
silent frames and signal runs is not considered here, but can certainly
improve results for sparse material. Replacing Rice coding with 
arithmetic coding \cite{arithmetic}, or hybrid entropy coding \cite{MPEG-ALS} 
gains typically few percent in compression efficiency.
We also experimented with an additional long-term 
predictor that tries to find the best matching segment to the current frame 
from the full history of the signal \cite{MPEG-ALS}, 
but did not include it in our models.

\subsection{Multichannel modeling}
\label{sec:multichannelmodeling}

With multichannel, or immersive audio, the  
question becomes whether correlations between channels can be exploited. 
In the case that $s$ has more than one channel, 
one can use model \ref{model:sep} for each channel $c$ separately; 
each sample $s_c(t)$ only is predicted from the past samples of that same channel.
The number of model parameters is then $pC$, where $C$ is the number of channels.
In contrast, we also construct a vanilla baseline for a multichannel 
predictor in order to test the hypothesis that there are easily 
exploitable correlations between the channels:

\begin{equation}\label{model:joint}
    s_c(t) = \sum_{c=1}^C\sum_{k=1}^p{\beta_{c,k} s_c(t-k)} + e_c(t).
\end{equation}

\noindent In effect, the current sample of each channel is predicted 
using all other channels' samples looking $p$ timesteps in the past.
This increases the amount of model parameters to $pC^2$.

Another possibility for exploiting the correlations between the channels 
is to utilize a \emph{transform} with desirable properties. For example, common 
technique is to use PCA or SVD to find a linear 
projection of maximal energy compaction, and orthogonality of the transformed 
components \cite{pcamultich, nimasvd}.
To our knowledge, this technique has not been well investigated for lossless 
multichannel audio coding, and it is only approximated with heuristic 
mid-side channel pairs etc. 
The SVD projection method of \cite{nimasvd} 
was utilized here.
To avoid large values, we found that applying the projection to the prediction 
residual $e$ and not to the original signal is preferable.
It should be notes that while computationally complex at the encoder, the 
more crucial decoding cost of such transform is only increased by a single matrix multiplication.

\subsection{Hierarchical reconstruction from downmix}

The main contribution of this paper is to suggest a multichannel audio 
signal model, which when optimized, can be used for efficient prediction 
in the context of hierarchical reconstruction.
Assume that the decoder has available many mixes of the same content, either
from the same container, or otherwise.
Decoding is traditionally done first on the lowest mix in the hierarchy 
(aka "downmix", typically the mix with the least amount of channels). 
It is then used to predict the next mix (aka "upmix") with the signal model.
The whole process can be repeated by using the decoded upmix as the new downmix 
for the next iteration.

We test adding simple additional predictors that utilize the downmix 
to the previous single-, and multichannel models of 
\eqref{model:sep} and \eqref{model:joint}:

\begin{equation}\label{model:sepdmx}
    s(t) = \sum_{k=1}^p{\beta_{k} s(t-k)} + 
    \sum_{d=1}^D{\gamma_d s_d(t)} + e_c(t),
\end{equation}

\begin{equation}\label{model:dmx}
    s_c(t) = \sum_{c=1}^C\sum_{k=1}^p{\beta_{c,k} s_c(t-k)} + 
    \sum_{d=1}^D{\gamma_d s_d(t)} + e_c(t),
\end{equation}

\noindent where $s_d$ indicates channel $d$ of the downmix, and 
$\gamma_d$ the corresponding prediction parameter.

It can be seen that these models only utilize the most current sample of the
downmix, in addition to predicting from the past of the upmix as in 
traditional models. We found this worked the best for our tests, as compared to more 
elaborate utilization of the downmix.
Also, the addition of such downmix prediction only introduces $D*C$ more model 
parameters.
Also important for the hierarchical models is the optimizer selection, 
as discussed in Sec.\ \ref{sec:modeloptimization}.

\subsection{Model optimization}
\label{sec:modeloptimization}

Traditionally, lossless coding predictors have been optimized with 
Levinson recursion \cite{levinson, burg}.
These methods achieve computational efficiency by assuming Toeplitz systems.
The Toeplitz assumption however limits the type of
predictors that are possible: all solved predictor parameters must 
originate from a time series of consecutive samples. Another alternative used by 
e.g. \cite{MPEG-ALS} is to use several cascaded Toeplitz models 
whose parameters are not optimized globally. In matrix form, 
the minimization becomes:

\begin{equation}\label{eq:toeplitz}
   \underset{\alpha} {argmin}||s - S'\alpha||_2^2
\end{equation}

\noindent where prediction source $S'$ is a Toeplitz matrix with 
different lags of source signal $s'$ as columns.
In case of standard single-channel model \ref{model:sep} of order $p$:

\begin{equation}
    \alpha = [\beta_1, \hdots, \beta_p].
\end{equation}



In contrast, we use well-established solvers for linear systems that 
are not limited to be Toeplitz, namely the GELSD algorithm 
available in LAPACK \cite{lapack}. Despite being slower computationally, 
it allows optimizing all model parameters globally when using 
the complicated models of \ref{model:joint} and \ref{model:dmx}, 
and include arbitrary columns to the source signal matrix $S'$.
When the predictor is based on the model of \ref{model:dmx}, we have:

\begin{equation}
    \alpha = [\beta_{1,1}, \hdots, \beta_{c,p}, \gamma_1, \hdots, \gamma_d],
\end{equation}

\noindent for prediction order $p$, $c$ upmix channels, and $d$ downmix
channels, respectively. 

To enable comparison, GELSD is used for all prediction models.
Computational efficiency refinements are largely left for future work.
We however utilize Tikhonov regularization \cite{tikhonov} in all solvers, 
except the single-channel baseline \ref{model:sep}, by solving for 
smaller (covariance) matrices, and adding a diagonal component $\delta I$:

\begin{equation}\label{eq:reg}
   \underset{\alpha} {argmin}||S'^Ts - (S'^TS' + \delta I)\alpha||_2^2.
\end{equation}

\noindent As is typical, the 16-bit integer input signals are transformed into 
double precision float in the range $[-1, 1]$ for optimization computations, 
For simplicity, all solved model parameters are quantized as 16-bit floats 
prior to the residual calculation and Rice coding. 
Mirroring the datatype changes and rounding operations in the decoder ensures 
lossless reconstruction.

\section{EXPERIMENTS}
\label{sec:experiments}

\subsection{Dataset}

We tested the methods for 100 songs that had been mixed and mastered specifically
to the 5.1. format. The content included mainly pop/rock, and classical genres from 
various different performers. 
In our view, such an ad-hoc dataset represents a general, realistic situation, 
and the exact content or music style is not a determining 
factor for the overall performance.
All material was utilized with 16 bit depth and 44.1 kHz sample rate in the tests.

LFE channel is omitted in the dataset; we only use the 5.0-channels (L, R, C, Ls, Rs) 
of the mixes. Preliminary experiments indicated that
including LFE in the prediction would not help, and thus sending it 
with a single-channel predictor like \eqref{model:sep} would just add the same 
amount of bitrate for all the methods in the comparisons.
Furthermore, LFE channel coding may benefit from advanced silence handling, 
which was not the focus of this paper.

We utilize the ITU standard downmix \cite{itumix} from 
5.0 to 2.0 stereo in order to show the benefit 
of hierarchical reconstruction in the typical situation where the 
downmix is correlated to the upmix. Of course, this is an artificial 
situation; in real life this downmix could be obtained at the decoder 
without sending it, by applying the known linear operation of \cite{itumix}.
However, we believe the results also indicate that there 
is a benefit when using an artistic downmix, 
especially if the processing in mixing consists of 
linear operations such as panning. 
This assumption may break down in rarer cases of strong nonlinearities 
or uncorrelated mixes.
It should also be emphasized that we are not in this paper addressing 
object audio, but channel-based material. The former aims to be agnostic to
the rendering setup by sending panning information per object, and thus can 
in principle account for the upmixing blindly.

\subsection{Systems tested}

The tested methods are listed in Table \ref{table:res}.
The models used for prediction discussed in Sec.\ \ref{sec:methods} had their 
parameters optimized with the Tikhonov regularized ($\delta = 1e-4$) GELSD solver 
for \eqref{eq:reg}. For the basic 
single-channel model of \eqref{model:sep}, GESLD was used to optimize 
\eqref{eq:toeplitz} in order to compare this baseline against FLAC with 
similar solver criterion. 
We used FLAC with the default parameters, as experimenting with other options 
resulted in little difference.

The prediction order for all models implemented 
(as well as the default FLAC LPC max order) was $p=8$.
Unlike in real-life coders, we used constant $p$ for each frame of 4096 samples. 
The only hyperparameter signaled per frame was the Rice code 
parameter per channel.
SVD projection of \cite{nimasvd} 
prior to residual coding was also applied selectively. 
In addition to quantizing the residual, 
the prediction and transform 
parameters were counted towards the 
bitrate of each method as discussed in Sec.\ \ref{sec:modeloptimization}.

As mentioned in Sec.\ \ref{sec:modeloptimization}, MPEG-ALS 
includes a more involved multichannel prediction models.
The reasons for not comparing against it here are the lack of availability of MPEG-ALS 
software, and the related fact that FLAC is more widely adopted.
See Sec.\ \ref{sec:results} for further discussion.

\subsection{Results}
\label{sec:results}

Table \ref{table:res} shows the average compression ratios 
for the 100 songs tested. Ratio per file was calculated as 
the total number of bits divided by original number of bits of the
16-bit representation. 
Even though the hierarchical systems rely on sending both 2.0 and 5.0 
mixes simultaneously, the 5.0 upmix compression ratio is more interesting
for evaluating the model effect. Since the 2.0-mix does not use hierarchical 
prediction, nor was the use of complex non-hierarchical models found to benefit, 
it can be sent with established stereo coding, such as FLAC in this experiment.
This merely adds the same constant rate for all tested methods 
when sending both 2.0 and 5.0.

\begin{table}[]
\centering
\begin{tabular}{ |p{5cm}|p{1cm}|p{1cm}| }
\hline
Upmix compression method & Total & Upmix \\
\hline
\hline
FLAC   & .545 & .540 \\
\hline
Single-channel, \eqref{model:sep} & .546 & .540 \\
\hline
Multi-channel, \eqref{model:joint} & .544 & .538 \\
\hline
Multi-channel, \eqref{model:joint} + SVD & .521 & \bf{.505} \\
\hline
*Single-channel + dmx, \eqref{model:sepdmx} + SVD & .527 & .515 \\
\hline
*Multi-channel + dmx, \eqref{model:dmx} & .487 & .458 \\
\hline
*Multi-channel + dmx, \eqref{model:dmx} + SVD & \bf{.387} & .318 \\
\hline
\end{tabular}
\caption{Average compression ratios of various models
(smaller is better). Middle column gives total ratio for sending both 
mixes (2.0 and 5.0), right column shows the effect for upmix (5.0) alone.
Methods indicated with * need the downmix available for decoding the upmix.}
\label{table:res}
\end{table}

It can be seen that in comparison to FLAC with the same database, the use of the 
baseline single-channel prediction model of \eqref{model:sep} 
gives close to identical performance. The vanilla multichannel model
(\eqref{model:joint}) does not give notable gains. 
However, when combined with subsequent residual SVD projection, the compression 
is improved. 
For the hierarchical methods requiring the presence of the downmix, and 
using it in prediction, the single-channel method of \eqref{model:sepdmx}
seems to not work well. The real benefit of using the downmix emerges 
when using the multichannel model of \eqref{model:dmx}, especially
when combined with the SVD projection. 
This implies that global parameter optimization can be an important 
factor for the success of complex signal model predictors.

Despite a direct comparison against real codecs not being the priority, 
it should be noted that compression ratios better than ours 
(or the present FLAC result) were reported 
for MPEG-ALS with a 5.1 test set \cite{MPEG-ALS}. 
However, the content in \cite{MPEG-ALS} may have been sparser and more dynamic, 
with less surround- and center channel utilization (material being older), 
and the LFE channel being included. 
Most importantly, the MPEG-ALS use of cascaded multichannel prediction
did not have as significant a relative benefit compared to 
non-multichannel baseline, as the 
utilization of the downmix prediction, or SVD projection in this paper.
Rather, it was comparable to the difference between our single- and multichannel 
baseline models, \eqref{model:sep} and \eqref{model:joint}.

\section{CONCLUSION}
\label{sec:discussion}

The work presents improved methods for the lossless compression of multichannel audio, both with an upmix alone and when the upmix is packed with a downmix, at the cost of computational complexity. Results show approximately a 30\% improvement in the compression ratio, over FLAC, when both the downmix and upmix are to be joinly encoded. A 10\% gain in compression ratio is achieved over FLAC, by utilizing a combination of multichannel prediction, SVD, and Rice coding, when sending 5.0-content alone. The proposed approaches could yield significant gains for data server storage and transmission of multichannel audio data. 
Further implementing frame-based method switching, silence handing, and other typical codec features may improve results for specific content.

\newpage

\bibliographystyle{IEEEbib}
\bibliography{strings,refs}

\end{document}